# MODIFICATION OF ROBERTS' THEORY FOR ROCKET EXHAUST PLUMES ERODING LUNAR SOIL

Philip T. Metzger
*NASA/KSC Granular Mechanics and Surface Systems Laboratory, KT-D3,*
*Kennedy Space Center, FL 32899, USA*
Philip.T.Metzger@nasa.gov

John E. Lane, Christopher D. Immer
*ASRC Aerospace, ASRC-15, Kennedy Space Center, FL 32899*
john.lane@ksc.nasa.gov, christopher.immer@ksc.nasa.gov

## Abstract

Roberts' model of lunar soil erosion beneath a landing rocket has been updated in several ways to predict the effects of future lunar landings. The model predicts, among other things, the number of divots that would result on surrounding hardware due to the impact of high velocity particulates, the amount and depth of surface material removed, the volume of ejected soil, its velocity, and the distance the particles travel on the Moon. The results are compared against measured results from the Apollo program and predictions are made for mitigating the spray around a future lunar outpost.

## Introduction

In preparation for the Apollo program, Leonard Roberts (1963) developed a remarkable analytical theory that predicts the blowing of lunar soil and dust beneath a rocket exhaust plume. Roberts assumed that the erosion rate is determined by the "excess shear stress" in the gas (the amount of shear stress greater than what causes grains to roll). The acceleration of particles to their final velocity in the gas consumed a portion of the shear stress. The erosion rate continues to increase until the excess shear stress is exactly consumed, thus determining the erosion rate. The shear stress balance equation is

$$\frac{1}{2} a u \frac{\mathrm{d}m}{\mathrm{d}t} = \left(\tau^* - \tau_0\right) = \Delta\tau \tag{1}$$

Where $\tau^*$ is shear stress in the gas, $\tau_0$ is shear strength of the soil, $u$ is gas velocity, $m$ is mass eroding per unit area of the regolith, and the derivative with respect to time $t$ is the mass erosion rate per unit area. The velocity the particle achieves in the gas is the product $au$, where the unitless parameter $a$ is the fraction of the gas velocity that the particle achieves. Roberts calculated it as a function of particle diameter $D$ and a number of other parameters as

$$a = \left[\frac{1}{2} + \sqrt{\frac{1}{4} + \frac{1}{\zeta}}\right]^{-1} \tag{2}$$

where

$$\zeta = \frac{18\mu_C h}{\sigma\sqrt{RT_C\left(4+k_{\text{hyper}}\right)}}\left[\frac{1}{D^2}+\frac{1}{D}\cdot\frac{\left(4+k_{\text{hyper}}\right)C_D}{72e\sqrt{2RT_C}}\frac{F_{\text{Thrust}}}{\mu_C h^2}\right] \tag{3}$$

where $\mu_C$ and $T_C$ are the gas viscosity and temperature in the combustion chamber, $R$ is the gas constant, $F_{\text{Thrust}}$ is the lander's thrust (assuming a single-engine lander), $h$ is the lander height, $\sigma$ is the particle density, $C_D$ is the coefficient of drag for a particle, $e$ is the base of the natural logarithm, and $k_{\text{hyper}}$ is the hypersonic parameter of the engine,

$$k_{\text{hyper}} = \gamma(\gamma-1)M_n{}^2 \tag{4}$$

where $\gamma$ is the ratio of specific heats of the gas and $M_n$ is the exit plane Mach number of the engine.

Roberts also calculated the largest $D_{\text{Max}}$ and smallest $D_{\text{Min}}$ particles that could be eroded based on forces at the particle scale, but the erosion rate equation assumes that only one particle size exists in the soil. He assumed that particle ejection angles are determined entirely by the shape of the terrain, which acts like a ballistic ramp, the particle aerodynamics being negligible. The predicted erosion rate and particle upper size limit appeared to be within an order of magnitude of small-scale terrestrial experiments, but Hutton (1968) showed that the predicted versus measured erosion rates had generally poor correlation. The lower particle size limit and ejection angle predictions were not tested.

## Modifications to Roberts' Theory

We have modified Roberts' theory in several ways. First we have observed in the Apollo landing videos that the ejection angles of particles streaming out from individual craters are time-varying and correlated to the Lunar Module (LM) thrust, thus implying that particle aerodynamics dominate in determining the ejection angle (Immer et al, 2008). We do not believe that there is sufficient understanding, yet, of these aerodynamic forces in the highly rarefied, high-Mach regime of lunar landings to be able to predict the ejection angles from first principles. This is discussed in the companion paper by Lane, et al, in these proceedings. Fortunately, we have found a method to measure the ejection angles at a number of points in the various Apollo landings, using the distortion of the LM's shadow to obtain information about the shape of the blowing dust cloud. This is explained in another companion paper (Immer et al, 2008) in these proceedings. So our first modification to Roberts' theory is to impose these ejection angles as measured in the Apollo landing videos. This is in lieu of a more sophisticated method being developed.

Second, we have integrated Roberts' equations over the lunar particle size distribution $P(D)$ (a probability density) to put in more realism about the nature of the soil. $P(D)$ tells number (count) of particles per unit particle diameter rather than the mass of soil per unit particle diameter. $P(D)$ was measured for JSC-1A lunar soil simulant using Sci-Tec's fine particle analyzer, as shown in Figs. 1 and 2.

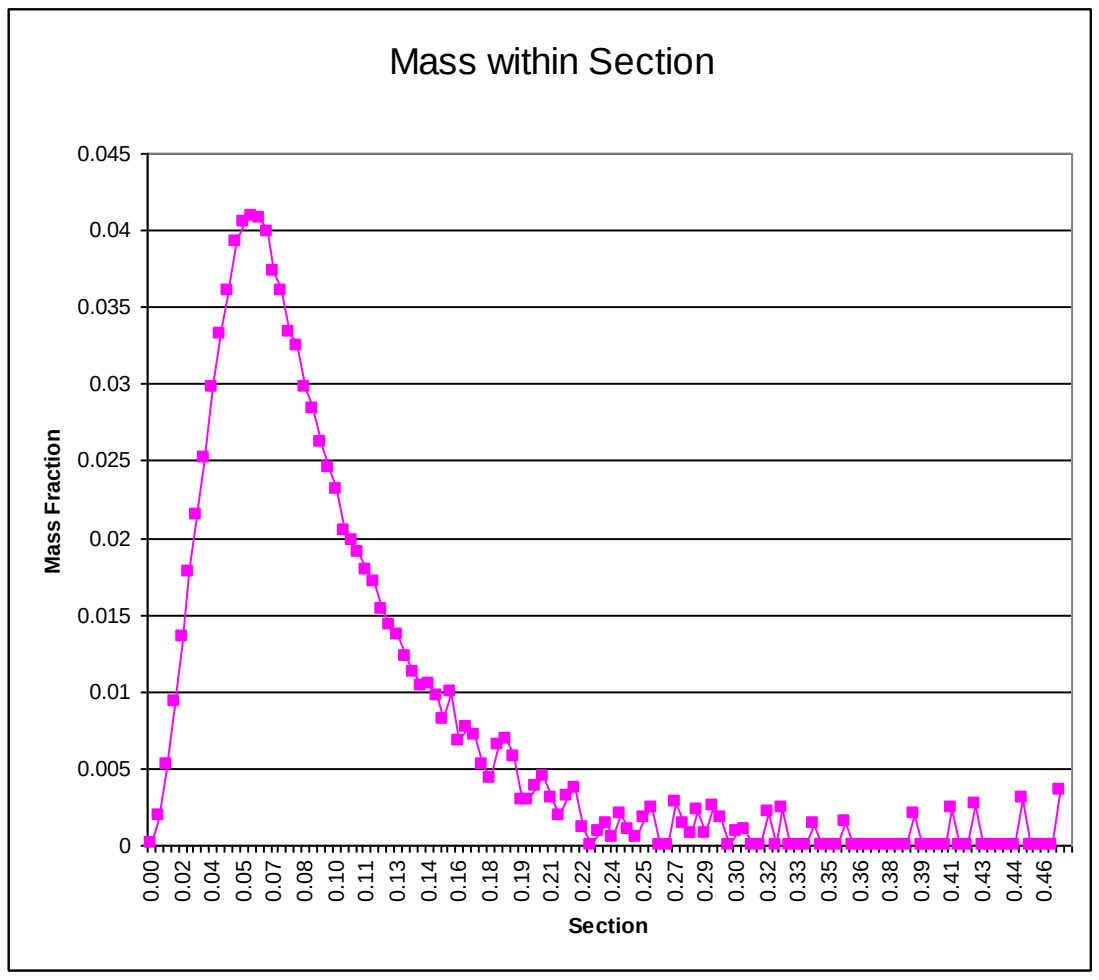

**Figure 1. Differential Mass Distribution for JSC-1A showing quantity of mass at each particle size, peaked near 60 microns**

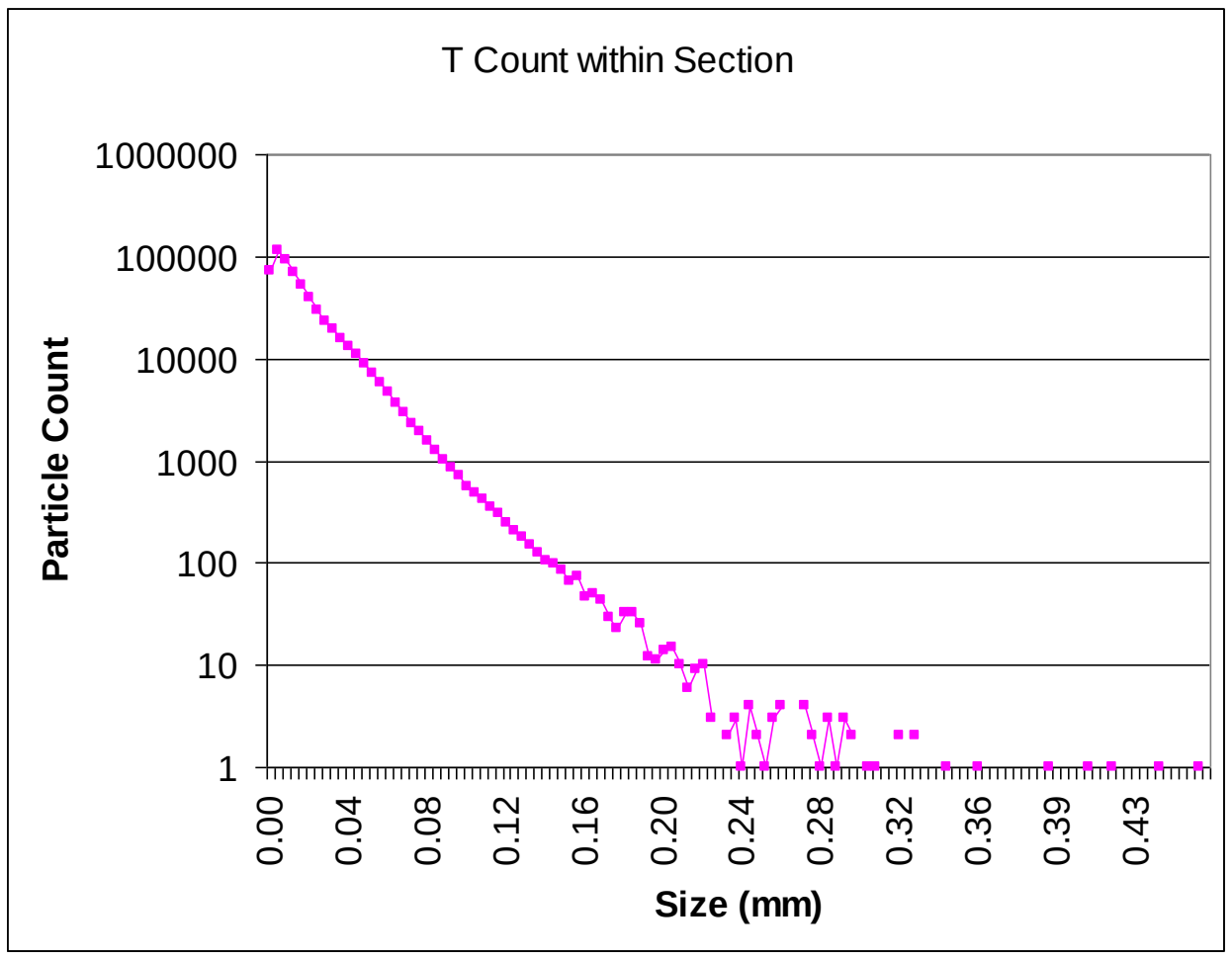

**Figure 2. Particle Size Distribution for JSC-1A showing number of particles at each particle size, peaked near 10 microns**

We find that the number count distribution is fit well by the function

$$P(D) \approx P_0 \exp\left(-\Lambda D^{\alpha}\right) \tag{5}$$

for all particle sizes greater than about 4 microns using values of $\Lambda$=36.6 mm$^{-1}$, $\alpha$=0.81, and $P_0$ chosen for normalization when integrating over ($D_{\text{Min}}$, $D_{\text{Max}}$). It over-predicts the number of particles we measured in JSC-1A for the single data point below 4 microns, but this is the region where JSC-1A may not accurately match the real lunar size distribution and where our experimental method may possibly have undercounted, and so Eq. 5 appears to be adequate until better data exist for that region. Introducing compact notation for the integral over $P(D)$,

$$\langle X \rangle = \int_{D_{Min}}^{D_{Max}} P(D) X(D) \, \mathrm{d}D, \qquad \hat{Y} = \frac{\langle D^3 Y \rangle}{\langle D^3 \rangle} \tag{6}$$

we obtain with some manipulation of Eq. 1,

$$\frac{\mathrm{d}n}{\mathrm{d}t} = \frac{12\Delta\tau}{\pi\sigma u \hat{a}} \tag{7}$$

where $n$ is the number of particles (per unit diameter) eroded per unit area on the lunar regolith, and where we have treated the particles as spheres to calculate their mass.

Third, we have modified Roberts' theory by adding a material damage model that predicts the number and size of divots that the impinging particles will cause in hardware surrounding the landing rocket. The volume of a divot is calculated using the equation of Sheldon and Kanhere (1972),

$$\tilde{V} = K_D v^3 D^3 \sigma^{3/2} H_V^{-3/2} \tag{8}$$

where $K_D$ is the ratio of vertical to horizontal force upon impact of the particle on a surface (assumed here to be unity), $v$ is the particle impact velocity, and $H_V$ is the Vickers hardness. Assuming the divot is a hemisphere we obtain its depth and area from this volume. The model is completed by using lunar gravity and particle ballistics to calculate the density of particles and their velocity as a function of particle size and location around the landing site. The model is integrated over a landing descent trajectory in order to calculate damage to hardware as a function of location around the landing site.

## Numerical Results

We have compared the predictions of this model with the measured surface damage to the Surveyor III spacecraft, which was subjected to the soil and dust sprayed by the landing of the Apollo 12 LM. Telemetry data from the Apollo LM landings is difficult to obtain at the present due to physical degradation of the tapes and the effort required to interpret the data formats to recover specific measurements out of the bit patterns. For now, we have obtained the approximate descent trajectory from the Apollo 12 LM by using the crew's voice callouts of altitude from the landing audio tape. The reconstructed (approximate) trajectory is shown in Figure 3.

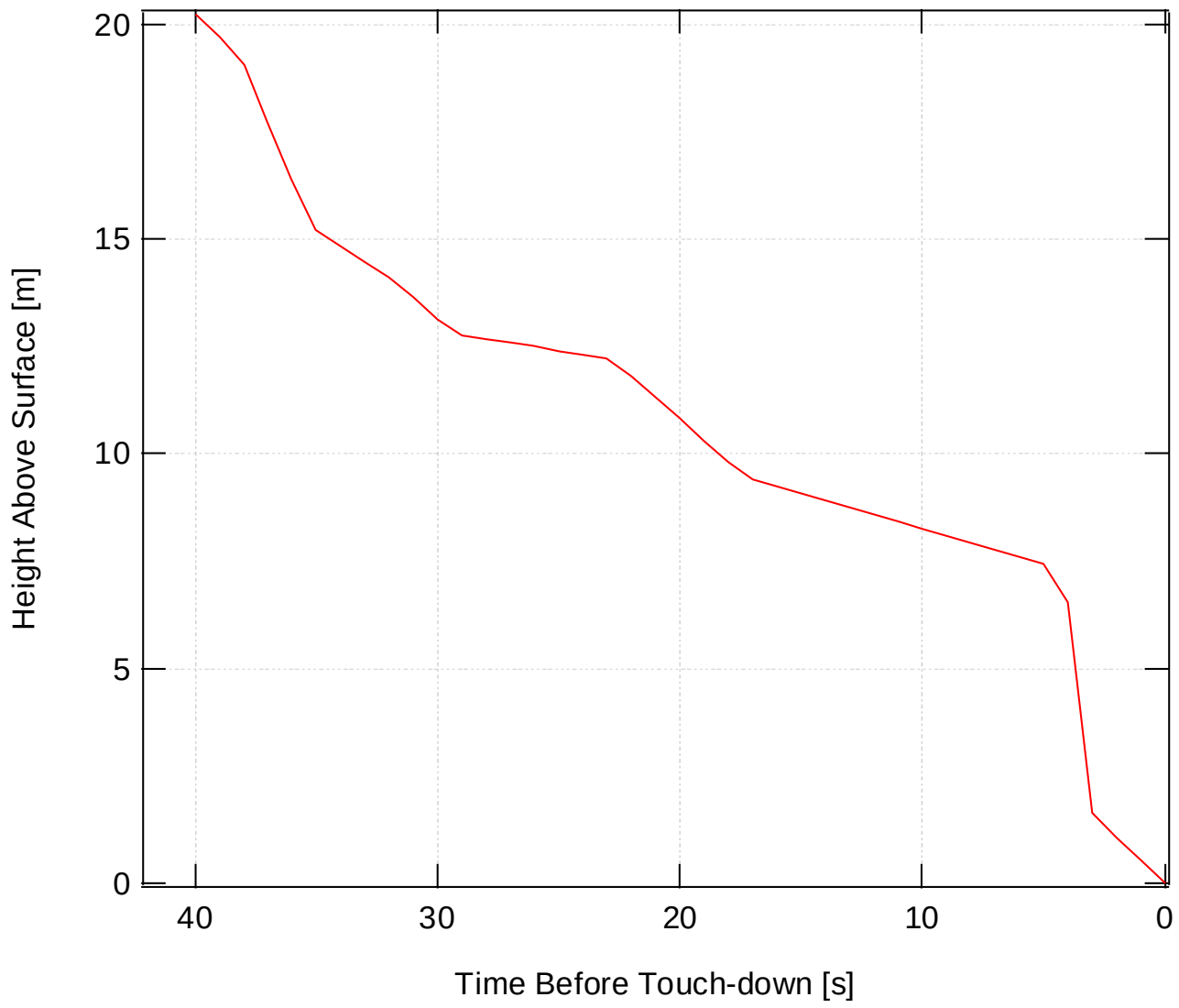

**Figure 3. Apollo 12 descent trajectory reconstructed from voice tapes.**

An example of the three dimensional output from the model is shown in Figure 4. The particle flux is shown in Figure 5.

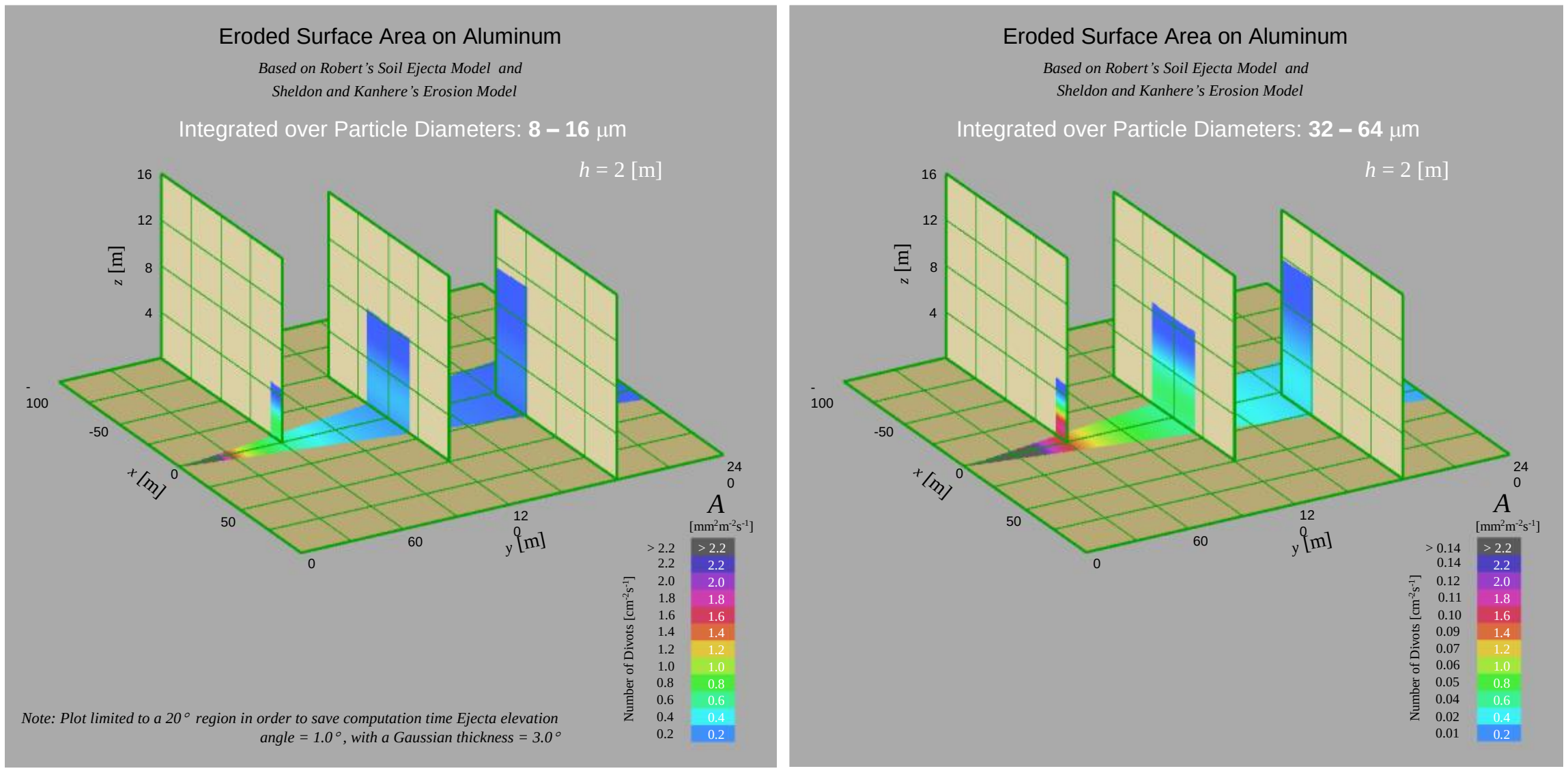

**Figure 4. Examples of three dimensional output from model, showing damage resulting from two different size ranges of lunar soil particles. Top: 8-16 microns. Bottom: 32-64 microns.**

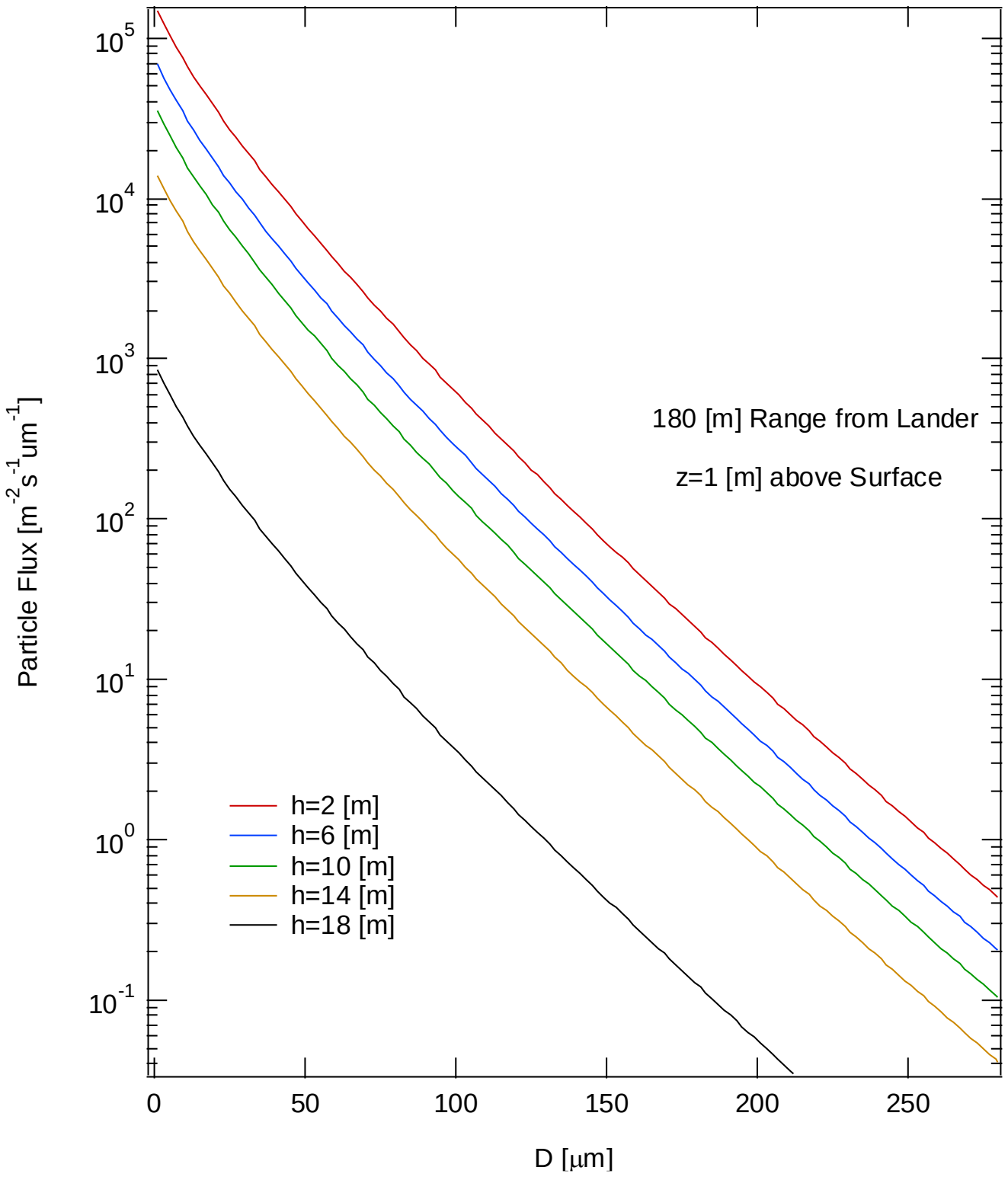

**Figure 5. Particle flux versus particle size for various lander heights.**

The particle velocity as a function of particle size is shown for several lander heights in Figure 6. The density of divots appearing per second on the surface of the Surveyor hardware is shown in Figure 7 as a function of lander height.

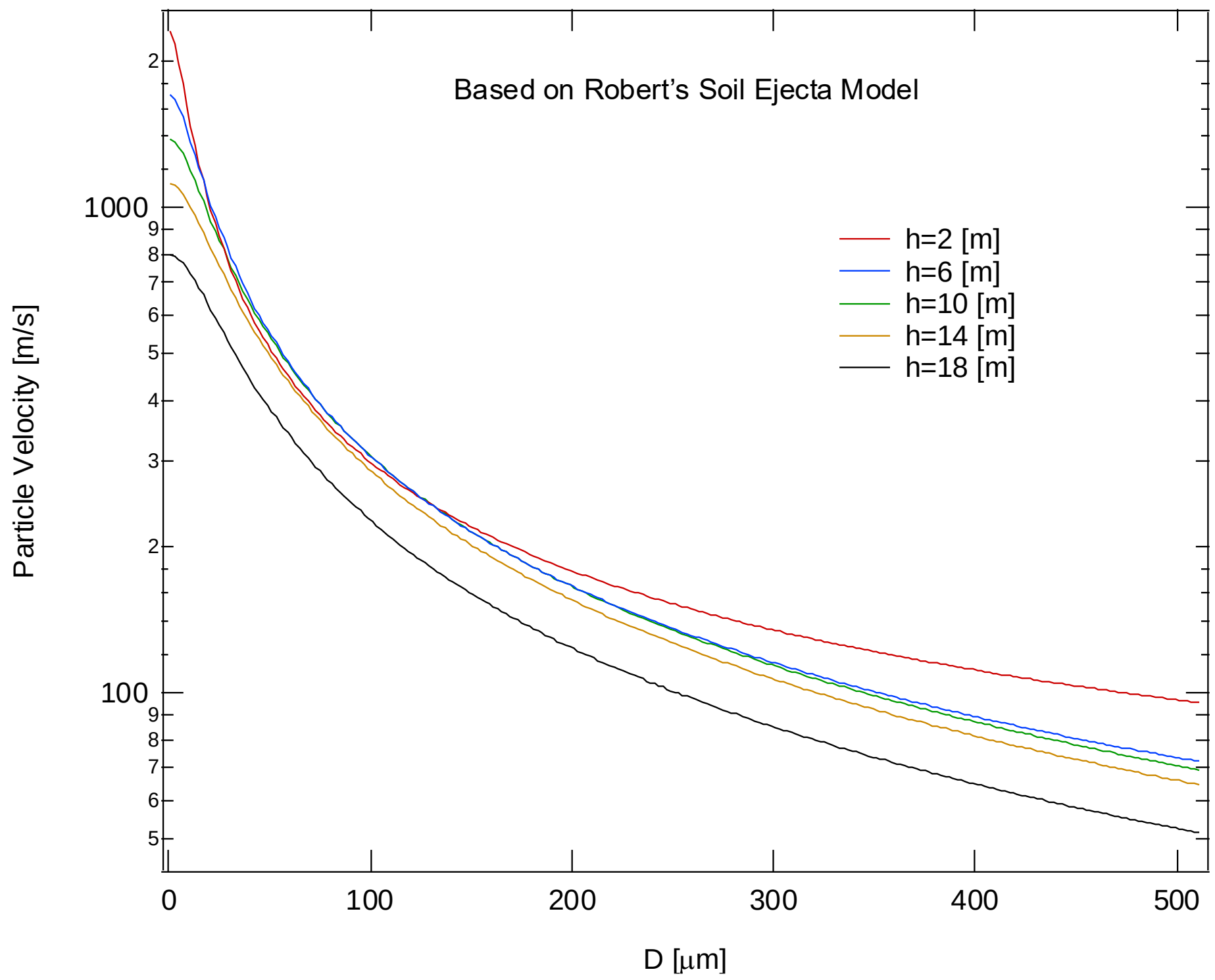

**Figure 6. Particle velocity as a function of size and lander height.**

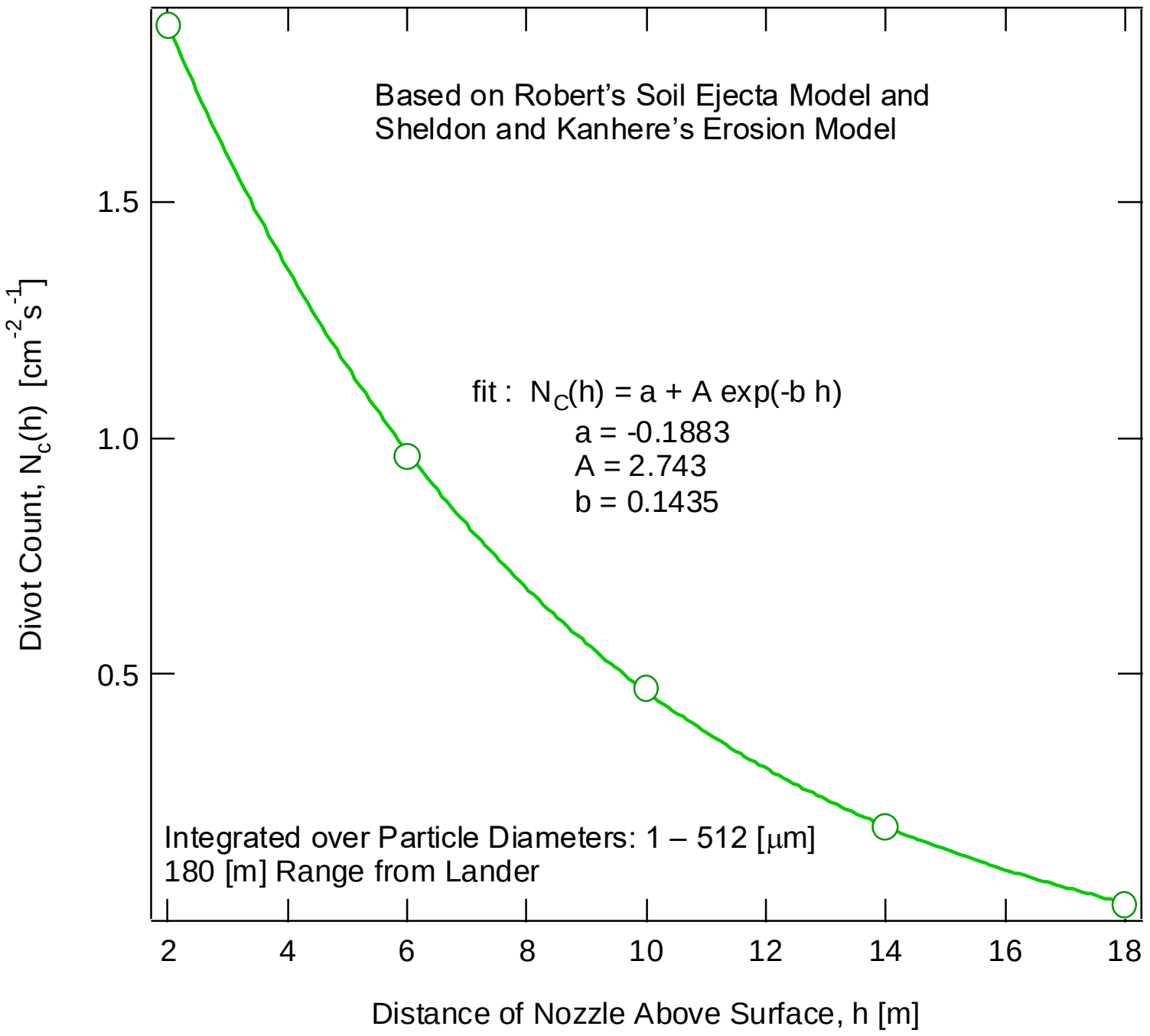

**Figure 7. Divots appearing per second per square centimeter on the Surveyor as a function of lander height.**

Integrating across the Apollo 12 LM trajectory, the analysis indicates that 3 divots/cm$^2$ should have resulted on the Surveyor hardware. This compares very favorably with the observed divots. We have estimated by reviewing several boxes of engineering logbooks (stored at the lunar curation facility at NASA's Johnson Space Center) that there were on the order of 1.4 divots /cm$^2$ on some portions of the hardware facing the descending LM. While the model's predicted order of magnitude is correct, we must point out that there are many sources of error in this analysis and the agreement must be understood to be largely coincidental at this stage of the modeling. For example, we do not know exactly the slope of the local terrain upon which the LM landed, and what the elevation angle was to the Surveyor relative to that local terrain slope. This and several other parameters are highly sensitive in determining the number of divots. Also, it should be noted that Immer et al (2008) have calculated the optical density of blowing lunar soil in the Apollo landing videos and have shown that the number of suspended particles in the cloud is at least two orders of magnitude higher than this model predicts. More work is needed to resolve the very large discrepancy.

The model also predicts the total volume of eroded soil to have been 787 liters. This is much higher than the prediction of 36-57 liters by Mason (1970). and somewhat less than the prediction of 1460-2080 liters derived from the calculations by Scott (1975) assuming he used a cylindrical crater geometry. Assuming the eroded soil was removed in an approximately conical shape out to the radius of erosion as predicted by the model, then the depth of the crater at the center would be only 1.3 cm. This would be unnoticeable impressed upon the much larger natural terrain features.

Finally, we note that the particle velocities predicted by the model approach lunar escape velocity. The ballistics of the particles after they leave the plume have been calculated and some examples are shown in Figure 8.

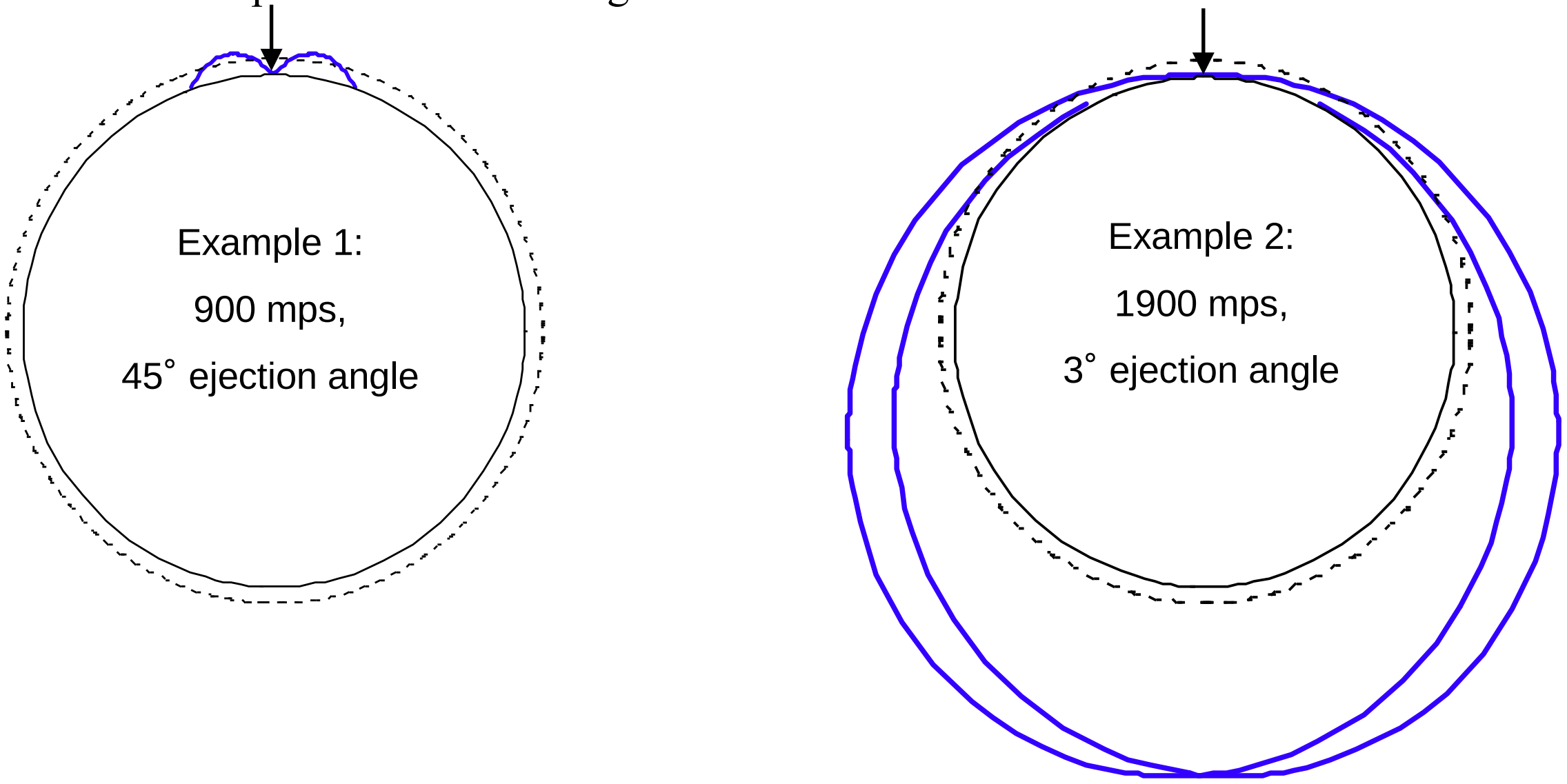

**Figure 8. Particle ballistics for two cases. (Solid black) lunar circumference. (Dashed) altitude of Command and Service Module orbit, for reference. (Arrow) landing site of LM. (Blue) ejected particle trajectory.**

## Conclusions

The modeling described above has made some progress in understanding the spray of soil in lunar landings, but primarily it has demonstrated what we do not know and shown us what some of the areas of future research should be. In particular, we need a better understanding of the erosion processes in this highly rarefied, supersonic flow regime so that we can better predict the erosion rate of soil. Some conclusions seem fairly certain, however. It seems certain that the particles are being ejected at very high velocity and therefore they are capable of causing surface erosion and pitting of surrounding hardware. The high erosion rate of the soil results in a cumulative effect upon the surrounding hardware that is significant and probably unacceptable for sensitive hardware. Also, because the particles travel at low elevation angle and at such high velocity, it is not feasible to land far enough away from other hardware to prevent unwanted effects from the soil spray. It therefore seems that we shall need to block the spray using physical barriers of some sort (berms, screens) or to prevent the spray altogether by stabilizing the soil at the landing site (lunar concrete, sintering, palliatives).

### Errata post-publication

p.1, *Roberts'* changed to *Roberts*.
p.2, inserted *(1968)* after *Hutton*.
p.2, corrected spelling of *simulant.*
p.4, corrected spelling of *degradation.*
p.8, added reference for *Hutton (1968).*